# Towards wafer scale inductive characterization of spin transfer torque critical current density of magnetic tunnel junction stacks

S. Sievers[1], N. Liebing[1], S. Serrano-Guisan[2], R. Ferreira[2], E. Paz[2], A. Caprile[3], A. Manzin[3], M. Pasquale[3], W. Skowroński[4], T. Stobiecki[4], K. Rott[5], G. Reiss[5], J. Langer[6], B. Ocker[6], H.W. Schumacher[1]

[1] Physikalisch-Technische Bundesanstalt, Braunschweig, Germany (e-mail: sibylle.sievers@ptb.de)

[2] International Iberian Nanotechnology Laboratory, Braga, Portugal

[3] Istituto Nazionale di Ricerca Metrologica, Torino, Italy

[4] AGH University of Science and Technology, Department of Electronics, Krakow, Poland

[5] Bielefeld University, Department of Physics, Bielefeld, Germany

[6] Singulus Technologies, Kahl am Main, 63796, Germany

Abstract

We explore the prospects of wafer scale inductive probing of the critical current density $j^{c0}$ for spin transfer torque switching of a CoFeB/MgO/CoFeB magnetic tunnel junction with varying MgO thickness. From inductive measurements magnetostatic parameters and the effective damping are derived and $j^{c0}$ is calculated based on spin transfer torque equations. The inductive values compare well to the values derived from current induced switching measurements on individual nanopillars. Using a wafer scale inductive probe head could in the future enable *wafer probe station* based metrology of $j^{c0}$.

*Terms*—Critical current density, spin transfer torque, MRAM

I. INTRODUCTION

Spin transfer torque (STT) is the basis of promising applications like STT oscillators, and magnetic random access memories (MRAM) [1]. One of the key material parameters for STT applications is the critical current density $j^{c0}$. It determines the conditions at which, depending on the device application, STT induced precessional self oscillations set in or at which current induced magnetization switching (CIMS) can be achieved. For MRAM applications this value is critical with respect to power consumption, drive circuit layout, storage density, write times, and the write/read threshold. It thus needs to be well optimized during material development and device design and tightly controlled during manufacturing. Today, determining $j^{c0}$ of a STT material such as a magnetic tunnel junction (MTJ) stack is a time consuming process. First, individual nanopillars with electrical contacts are fabricated from the MTJ stack by a multi mask high resolution clean room lithography process. Then the pillars are electrically contacted and CIMS experiments are carried out to determine $j^{c0}$. Formerly, two other key MRAM material parameters, the tunnel magneto resistance (TMR) and the resistance area (RA) product could also only be determined on individually contacted devices. However, later a lithography-free testing scheme using current in-plane tunneling [2] enabled fast material research and wafer scale testing [3] thereby boosting MRAM and MTJ sensor development. In analogue, the development of a lithography-free and wafer scale characterization scheme for $j^{c0}$ could underpin efficient STT material development and could allow fast wafer scale in-line testing for quality control.

Here we explore the prospects of wafer scale inductive probing of $j^{c0}$ of a typical MRAM material: a CoFeB/ MgO/ CoFeB based MTJ stack with varying MgO thickness $t_{MgO}$. From ferromagnetic resonance (FMR) measurements [4] of the unpatternd MTJ stacks we derive the magnetostatic parameters and the effective damping $\alpha$ as function of $t_{MgO}$ and calculate $j^{c0}$ using the spin transfer torque equations [5]. The derived values of $j^{c0}$ compare well to the values derived from current induced switching measurements on individual nanopillars of the same MTJ stack. We further compare the relevant parameters derived from inductive measurements using a standard coplanar waveguide [6] and using an inductive probe head suitable for wafer scale testing [7]. From the comparison we conclude that wafer scale inductive determination of $j^{c0}$ of STT materials seems possible.

Ferromagnetic resonance (FMR) measurements of magnetic thin films and multilayers can be carried out using either cavity based or coplanar waveguide based systems. The latter can be carried out in time domain using pulsed inductive microwave magnetometry (PIMM) [8] or in frequency domain by vector network analyzer (VNA) FMR [9]. The coplanar broad band techniques deliver the FMR frequency $f_{FMR}$ and the line with $\Delta f$ as function of magnetic field. Fitting $f_{FMR}$ to a Kittel model yields the magnetostatic material parameters (saturation magnetization $M_S$, anisotropy $K_i$, …) and also interlayer exchange coupling $J_{FL}$ between ferromagnetic layers. Analysis of $\Delta f$ yields the effective magnetization damping $\alpha_{eff}$. Note that according to Slonczewski's STT model $j^{c0}$ is directly proportional to $\alpha_{eff}$, $M_S$ and the effective anisotropy of the material [10,11,12]. Hence, one can calculate $j^{c0}$ based on the static and dynamic parameters derived from inductive measurements.

In the first part of this work we summarize the theory for the calculation of $j^{c0}$ based on the inductive characterization of an MTJ sample. In the second part $\alpha_{eff}$ and $M_S$ of a set of MTJ test structures with different MgO barrier thickness $t_{MgO}$ are determined from inductive measurements [4] and the corresponding critical current densities are calculated. The feasibility of the inductively derived $j^{c0}$ is confirmed by CIMS data obtained on patterned nanopillars from the same material stack [13]. In the last part we compare the inductive measurements of a further MTJ stack using two different inductive setups: a standard coplanar wave guide and a recently developed wafer scale inductive probe head [7].

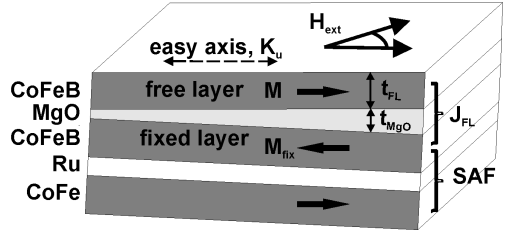

FIG. 1. Simplified sketch of the MTJ stack. The CoFe layer of the synthetic antiferromagnet (SAF) fixing the CoFeB layer is pinned to a PtMn antiferromagnet (not shown).

## II. THEORY

In our experiments we study an MTJ, which is modeled as a coupled two layer system consisting of a free layer (FL) with in-plane magnetization $M$, thickness $t_{FL}$ and saturation magnetization $M_S$ and a fixed layer as a reference layer with an in-plane magnetization $M_{fix}$. The FL has an in-plane uniaxial anisotropy $K_u$ and, at zero external field, the system has two stable magnetic states of the two electrodes, parallel (P) and antiparallel (AP). When currents are applied to the device, the magnetization dynamics of the FL can be described by the Landau-Lifshitz- Gilbert equation

$$\frac{dM}{dt} = -\gamma\mu_0 M \times H_{eff} - \frac{\alpha_{eff}}{M_s} M \times \frac{dM}{dt} + \tau_{STT}\frac{M\times(M\times M_{fix})}{M_S^2 M_{fix}} + \tau_{FLT}\frac{M\times M_{fix}}{M_S M_{fix}}, \quad (1)$$

including the Slonczewski or in-plane STT term and the field-like or perpendicular STT term, respectively. Here, $\alpha_{eff}$ is the effective damping and $\gamma$ the gyromagnetic ratio. According to this Sloncziewski's model $\tau_{STT}$ is a function of the current density $j$, $\tau_{STT} = \gamma\hbar\eta/(2et_{FL})j$, e is the absolute value of the electronic charge and $\eta$ is the spin-torque efficiency factor. The STT terms modify the magnetization dynamics of the device leading to magnetic excitations that can induce FL magnetization reversal. In zero external field the onset of such STT excitations is determined by $\tau_{STT}$ [14]:

$$\tau_{STT} = \alpha_{eff}\gamma M_s \left[\frac{2K_u}{M_s} + \frac{1}{2}\mu_0 M_s \pm \mu_0 H_{int}\right] = \alpha_{eff}\gamma M_s\mu_0[H_u + H_D \pm H_{int}] = \alpha_{eff}\gamma M_s\mu_0 H_{eff}^0. \quad (2)$$

$H_{int} = J_{FL}'/(t_{FL}\mu_0 M_S)$ is an interaction field taking into account both the magnetostatic and the exchange interaction $J_{FL}$ between FL and fixed layer. $H_u=2K_u/\mu_0 M_S$ is the uniaxial anisotropy field and $H_D=M_S/2$ is the demagnetizing field of the FL. They sum up to an effective field $H^0_{eff}=H_u+H_D\pm H_{int}$.

For symmetric tunnel junctions this leads to a critical current expression for magnetization switching from P to AP $j_{P \to AP}^{c0}$ and AP to P $j_{AP \to P}^{c0}$

$$j_{P \to AP/AP \to P}^{c0} = \mp 2 \frac{e}{\hbar} \frac{\alpha_{eff} t_{FL} \mu_0 M_S}{\eta} H_{eff}^0, \quad (3)$$

with $\eta = \frac{1}{2} P/(1+P^2 \cos \Theta)$ [13,15]. The spin polarization $P$ can be calculated from TMR values (using Julliere's model), which are accessible via non-destructive wafer scale measurements [3]. Thus, from the parameters $\alpha_{eff}$, $M_S$, $J_{FL}$ and $K_u$ we can calculate $j^{c0}$.

A determination of these parameters is possible from non-invasive inductive techniques as will be discussed for the example of VNA-FMR measurements. The resonance frequency $f_{FMR}$ is determined by the free energy density function $F$. In case of a uniaxial anisotropy $F$ is given by

$$F = -\mu_0 M_S H_{ext} m(\phi) h(\varphi) - \frac{\mu_0}{2} M_S^2 \cos^2(\theta) - K_u \sin(\theta) \cos^2(\phi) - \frac{J_{FL}}{t_{FL}} \cos(\phi), \quad (4)$$

where $\varphi$ and $\phi$ are the azimuthal (in-plane) coordinates of the FL magnetization and the external field $H_{ext}$, respectively, with respect to $M_{fix}$, and $\theta$ is the polar (out-of-plane) coordinate of the FL magnetization. From $F$, following the ansatz of Smit and Beljers [16] the field dependence of the precession frequency can be derived:

$$f_{FMR} = \frac{\gamma \mu_0}{2\pi} \sqrt{H_{ext} \cos(\varphi - \phi) + H_u \cos(2\phi) + H_{int} \cos(\phi)} \cdot \sqrt{H_{ext} \cos(\varphi - \phi) + H_u \cos^2(\phi) + H_{int} \cos(\phi) + M_S}. \quad (5)$$

By fitting this model to the measured field dependence of $f_{FMR}$ from the FMR measurements $H_u$, $H_{int}$ and $M_s$ can be found. Eq. 5 also holds for time domain (PIMM) measurements.

The last parameter inquired for the $j^{c0}$ calculation is the effective damping $\alpha_{eff}$. The effective damping emerges from the linewidth of the FMR peak. In case of small damping the imaginary part of the absorption peak can be fitted to a Lorentzian Function with the line width $\Delta f$:

$$\alpha_{eff} = \frac{4\pi}{\gamma \mu_0 M_S} \Delta f. \quad (6)$$

In summary from non invasive inductive measurements like PIMM or VNA-FMR we can deduce the intrinsic magnetic properties of the sample necessary to calculate $j^{c0}$.

## III. EXPERIMENTS: INDUCTIVE DETERMINATION OF $j^{c0}$

For the comparison of CIMS and inductive measurements we studied MTJ stacks with different thicknesses of the tunneling barrier on Si wafers with a layer sequence: Ta(5)/CuN(50)/Ta(3)/CuN(50)/Ta(3)/PtMn(16)/CoFe(2)/Ru(0.9)/Co$_{40}$Fe$_{40}$B$_{20}$(2.3)/-MgO($t_{MgO}$)/Co$_{40}$Fe$_{40}$B$_{20}$(2.3)/Ta(10)/CuN(30)/Ru(7) from bottom to top [17]. Numbers in parentheses refer to the layer thickness in nm. The MgO thickness is varied from $t_{MgO}$=0.62 to 0.96 nm without changing the remaining stack. The CoFe layer is the bottom part of a synthetic antiferromagnet (SAF) and is pinned by coupling to the antiferromagnetic PtMn.

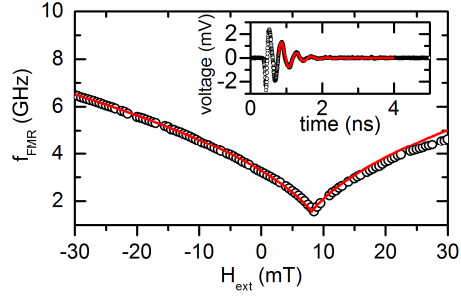

FIG. 2. Static field dependence of $f_{FMR}$ derived from PIMM for $t$MgO = 0.76 nm. Red line is the fit to Eq. (5). The inset shows the inductive PIMM data at 5mT easy axis field and the fit to a damped sinusoid (red).

CIMS measurements were performed on about five nanopillars for each $t_{MgO}$=(0.96, 0.88, 0.82 and 0.71) nm by applying voltage pulses of length 1 ms ≤ $\tau$ ≤ 54 ms. Details can be found elsewhere [13]. The $j^{c0}$ from CIMS serve as reference to validate the inductive measurements. The TMR and RA were determined by wafer scale measurements using a commercial current in-plane tunneling setup [2,3]. For thick MgO barriers ($t_{MgO}$≥0.75 nm) the TMR ratio is high (TMR>150%) and almost thickness independent. For thinner barriers ($t_{MgO}$<0.71 nm) it drops significantly, pointing to possible barrier imperfections.

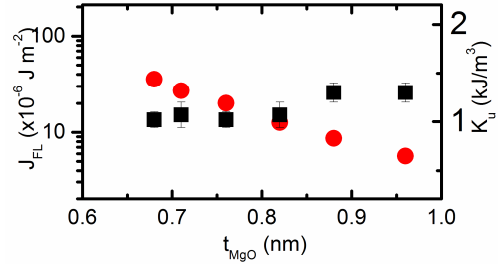

FIG. 3. (Color online) MgO thickness dependence of the coupling $J_{FL}$ between free and reference layer (red circles) and uniaxial anisotropy energy $K$u of the free layer (squares) from PIMM measurements.

For inductive characterization pieces of 2×4 mm$^2$ lateral dimension were cut from the MTJ wafer. PIMM measurements were performed at room temperature with easy axis external field. The MTJ stacks were placed on top of a coplanar waveguide (CPW) contacted with microwave probes. The setup can be used both for frequency domain and time domain inductive measurements. Details of this "standard" setup and the PIMM measurement technique are reported elsewhere [6,18]. From a single time resolved PIMM measurement at a given external field $H_{ext}$, the precession frequency and the damping parameter of the FL at this field value can be extracted by fitting to an exponentially damped sinusoid. The inset to Fig. 2 shows typical time resolved PIMM data for $t_{MgO}$=0.76 nm and $H_{ext}$=-5 mT. The measured field dependence of the FL precession frequency (open dots) was fitted by the model function in Eq. (5) (red line) to determine the FL magnetic parameters $M_S$, $K_u$ and $J_{FL}$ of the MTJ stacks. A constant magnetization saturation of $M_S$=1.35 T is obtained for all samples. The results for $J_{FL}$, $K_u$ and $\alpha_{eff}$ are summarized in Figs. 3 and 4.

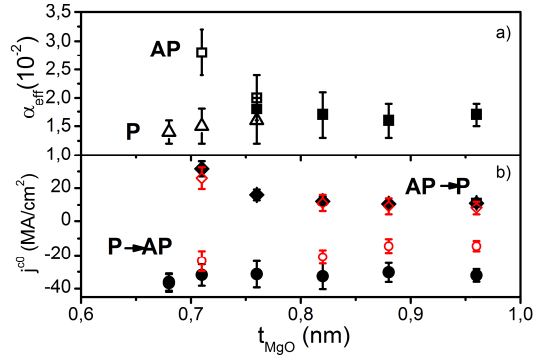

FIG. 4 (a): effective damping $\alpha_{eff}$ vs. MgO thickness. Open triangles refer to the damping parameter at parallel ($\alpha_P$), open squares to the damping parameter at antiparallel configurations ($\alpha_{AP}$). Full squares to equal $\alpha_{eff}$ of both configurations. b): Inductively determined $j^{c0}$ (black symbols) for P→AP ($j^{c0}_{P \to AP}$, circles) and AP→P ($j^{c0}_{AP \to P}$, rhombs) switching. Red symbols show $j^{c0}$ as determined from CIMS experiments on individual nanopillars.

In Fig. 3 $J_{FL}$ shows an exponential decrease with the barrier thickness for $t_{MgO} \geq 0.7$ nm. At low thicknesses a ferromagnetic coupling between the FL and the fixed layer is observed. In contrast $K_u$ shows no significant thickness dependence. In Fig. 4(a) three different regimes of $\alpha_{eff}$ can be observed, depending on $t_{MgO}$: for $t_{MgO}$>0.76 nm no influence of the FL magnetization orientation (AP,P) on the FL damping is found and an almost constant value of $\alpha_{eff} \sim 0.017 \pm 0.01$ is observed. For 0.68 nm<$t_{MgO}$ 0.76 nm a different $\alpha_{eff}$ is obtained in the AP/P state ($\alpha_{AP}/\alpha_P$ marked by open squares/triangles). The measured increased damping in AP state is attributed to *orange-peel* coupling at low $t_{MgO}$ due to barrier roughness [4]. At even lower barrier thickness FL precession is only observed in the P state and $\alpha_{AP}$ cannot be determined. Note that this thickness range below $t_{MgO}$=0.7 nm is not suitable for MRAM applications as no stable AP state can be realized. These inductively determined parameters are used to calculate the expected $j^{c0}$ as a function of $t_{MgO}$ as shown in Fig. 4 (b).

The black symbols mark the inductively determined values of $j^{c0}$ for AP→P (full rhombs) and P→AP (full circles). In the case of unequal effective damping for P, AP the according relevant values $\alpha_P$ (P→AP) or $\alpha_{AP}$ (AP→P) were used to determine $j^{c0}_{P\rightarrow AP}$ and $j^{c0}_{AP\rightarrow P}$. The red open symbols mark the values of $j^{c0}$ as determined from CIMS experiments on nanopillars fabricated from the same MTJ stack. For AP→P switching both data sets show an excellent agreement over the whole range of $t_{MgO}$ confirming the feasibility of inductive determination of $j^{c0}$. For P→AP reversal a deviation of the two values beyond the measurement uncertainty is found for $t_{MgO} > 0.85$ nm. Here the inductively determined $j^{c0}$ exceeds the CIMS value by about 50 %. The reason for the difference for AP→P and P→AP is not fully clear. It might be related to the influence of a field like torque or different resistive heating in the P and AP states during CIMS. Note, however, that for the lower thickness range of P→AP reversal the inductive and CIMS data of $j^{c0}$ agree within the measurement uncertainty. Note further that the general trends of the thickness dependence of $j^{C0}$ of both data sets agree well. For $t_{MgO} > 0.75$ nm $j^{C0}$ is almost independent of $t_{MgO}$. In contrast for lower $t_{MgO}$ the absolute value of $j^{C0}$ increases with decreasing $t_{MgO}$. Especially for AP→P this increase can be well explained by the increase of $\alpha_{AP}$ due to orange peel coupling via the thin MgO barrier. To summarize, the results of the inductively determined $j_{C0}$ show a good agreement with the CIPT measurements opening the path towards a future inductive and non-destructive determination of this key STT material parameter.

IV. INDUCTIVE PROBE HEAD MEASUREMENTS

We have previously described a CPW probe head (PH) suitable for wafer scale FMR [7]. The head consists of a CPW with rear contacts that can be brought in contact with the magnetic film on a wafer. So far it has been tested on single layer magnetic thin films. In the following we will compare FMR results of an MTJ stack using the wafer scale PH-FMR and our standard (S)-FMR as used in Sec. III.

We compare the inductive data obtained on a similar MTJ stack as described above, but with different thickness of the magnetic layers: Ta(5)/CuN(50)/ Ta(5)/CuN(50)/ Ta(5)/Ru(5)/ IrMn(20)/ CoFe(2)/ Ru(0.85)/ $Co_{40}Fe_{40}B_{20}$(2.6)/ MgO($t_{MgO}$)/ $Co_{40}Fe_{40}B_{20}$(2)/Ta(10)/Ru(7) [17]. This wafer was cut into pieces of (i) 20x20mm$^2$ (larger than the PH size) to test "wafer scale" measurements and (ii) 5x5 mm$^2$ size for characterization by S-VNA. VNA-FMR measurements were performed using both setups [18]. $f_{FMR}$ and $\Delta f$ of the frequency domain resonances peaks were calculated by fitting the sum of a symmetric and an antisymmetric Lorentzian to the respective data. $\alpha_{eff}$ was calculated from measured $\Delta f$ via Eq. 6.

A typical measurement result of $f_{FMR}$ (a) and $\alpha_{eff}$ (b) is shown in Fig. 5. The data is shown for negative applied fields $\mu_0 H$ and hence for AP configuration (full configuration of FL, fixed layer and bottom SAF layer is sketched). Note that our PH setup allows the application of higher fields and hence the wider PH data range. In the overlapping data range $f_{FMR}$ obtained by S-FMR

(red) and PH-FMR (black) agree well. Fitting $f_{FMR}$ to Eq. 5 yields a good agreement of magnetostatic parameters: PH-FMR: $\mu_0 M_S$=0.84±0.02 T, $J_{FL}$=13±1 µJ/m$^2$; S-FMR: $\mu_0 M_S$ =0.81±0.02 T, $J_{FL}$ =12±1 µJ/m$^2$) with $K_u$ set to 1200 J/m$^3$. Note that the present centimeter size PH in contact with the MTJ wafer reveals transmission discontinuities due to standing GHz waves inhibiting data analysis for certain frequencies (PH data gaps). When $f_{FMR}$ approaches these gaps the PH-FMR resonance data are no longer described by a Lorentzian and therefore, $\Delta f$ derived from PH-FMR and hence PH-$\alpha_{eff}$ are artificially enhanced. This effect can be seen in Fig. 5(b) where $\alpha_{eff}$ derived by both setups is plotted. The S-FMR damping (red) is almost constant over the given measurement range yielding an average effective damping of $\alpha_{eff}$ =0.047±0.003. The PH-FMR data shows a slightly lower damping in the center of the accessible data ranges and a strong increase of $\alpha_{eff}$ when $f_{FMR}$ approaches frequency gaps. Averaging over the displayed data yields $\alpha_{eff}$=0.04±0.01, a comparable value as for the S data but with larger uncertainty.

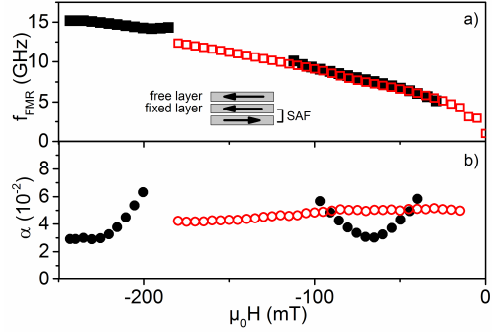

FIG. 5. (a) $f_{FMR}$ vs. $H$ measured by the inductive probe head (PH, black) and standard VNA-FMR (S, red open). Magnetic configuration is sketched. (b) $\alpha_{eff}$ as calculated from the linewidths of the resonance peaks from PH (black) and S (red open dots) measurements.

These results demonstrate the principal feasibility of a wafer scale FMR for the determination of key material parameters for STT MRAM. For future low uncertainty probing of $j^{C0}$ the probe heads need further optimization to avoid detrimental CPW resonances. A possible approach might be a bended CPW on a flexible substrate with reduced contact area to the stack and a larger distance to the high frequency connectors.

This work was supported by EMRP JRP IND 08 MetMags. The EMRP is jointly funded by the EMRP participating countries within EURAMET and the European Union. W.S. and T.S. acknowledge Polish National Science Center statutory grant 11.11.230.016.